\documentclass[
aps,pra,
a4paper,
reprint,
onecolumn,
12pt,
longbibliography,
nofootinbib,
preprintnumbers,
superscriptaddress
]{revtex4-2}
\usepackage[utf8]{inputenc}
\usepackage[english]{babel}
\usepackage[T1]{fontenc}

\usepackage{amsmath}
\usepackage{amssymb}
\usepackage{graphicx}

\usepackage{hyperref}
\hypersetup{citecolor=blue}
\hypersetup{colorlinks=true}
\hypersetup{linkcolor=blue}
\hypersetup{urlcolor=blue}

\newcommand*\sumprime{\mathop{\sum\nolimits'}}

\begin{document}

\title{On the simple derivation of the Casimir effect}
\author{Hai-Chau Nguyen}
\email{chau.nguyen@uni-siegen.de}
\affiliation{Naturwissenschaftlich--Technische Fakultät,
Universität Siegen, Walter-Flex-Straße 3, 57068 Siegen, Germany}

\author{Matthias Kleinmann}
\email{matthias.kleinmann@uni-muenster.de}
\affiliation{Department for Quantum Technology, Universität Münster, Heisenbergstraße 11, 48149 Münster, Germany}
\affiliation{Naturwissenschaftlich--Technische Fakultät,
Universität Siegen, Walter-Flex-Straße 3, 57068 Siegen, Germany}

\begin{abstract}
The Casimir effect in its simplest form describes the attraction of two parallel conducting plates at close distance due to the vacuum fluctuation of the electromagnetic field.
Its derivation can be found in many introductory works on quantum optics.
Here we return to the original paper by Casimir and find subtle nuances in his derivation that are worth discussing to give a complete picture of a mathematically sound derivation of the effect.
\end{abstract}

\maketitle
In his seminal paper \cite{Casimir1948}, Casimir showed that when one puts two large metal plates of size $L \times L$ at a small distance $a$ from each other, a negative force of
\begin{equation}
    F = - \frac{\pi^2 L^2}{240} \frac{\hbar c }{a^4}
\label{eq:Casimir-force}
\end{equation}
occurs as a consequence of the quantum fluctuations of the electrodynamic vacuum.
He demonstrated this through an elegant calculation within three pages.
This simple derivation has since then made its way as a pedagogical introduction to the topic in the literature \cite{Schleich2001a, Gerry2005a, Leonhardt2010a,Dutra2005a, Milonni1993a,Schwartz2013a}, also when other more advanced derivations are known \cite{Milonni1993a, Buhmann2013a, Jaffe2003Unnatural, Graham2003CasimirPLB, Graham2002VacuumNuclPhysB, Milton2004Casimir, Bordag2001New, Schwartz2013a}.
Unfortunately, when Casimir's argument is presented in condensed form, one step is sometimes {compressed} in a way that renders the derivation inconsistent, which can make it appear mysterious to students approaching the topic.
We should emphasize that this is independent of the discussions on different subtle aspects and more advanced derivations of the Casimir effect in the contemporary literature \cite{Milonni1993a, Buhmann2013a, Jaffe2003Unnatural, Graham2003CasimirPLB, Graham2002VacuumNuclPhysB, Milton2004Casimir, Bordag2001New}.
{In this article we} present this condensed argument and then reproduce Casimir's original argument \cite{Casimir1948} for a comparison. 

\begin{figure}
    \centering
    \includegraphics[width=0.6\textwidth]{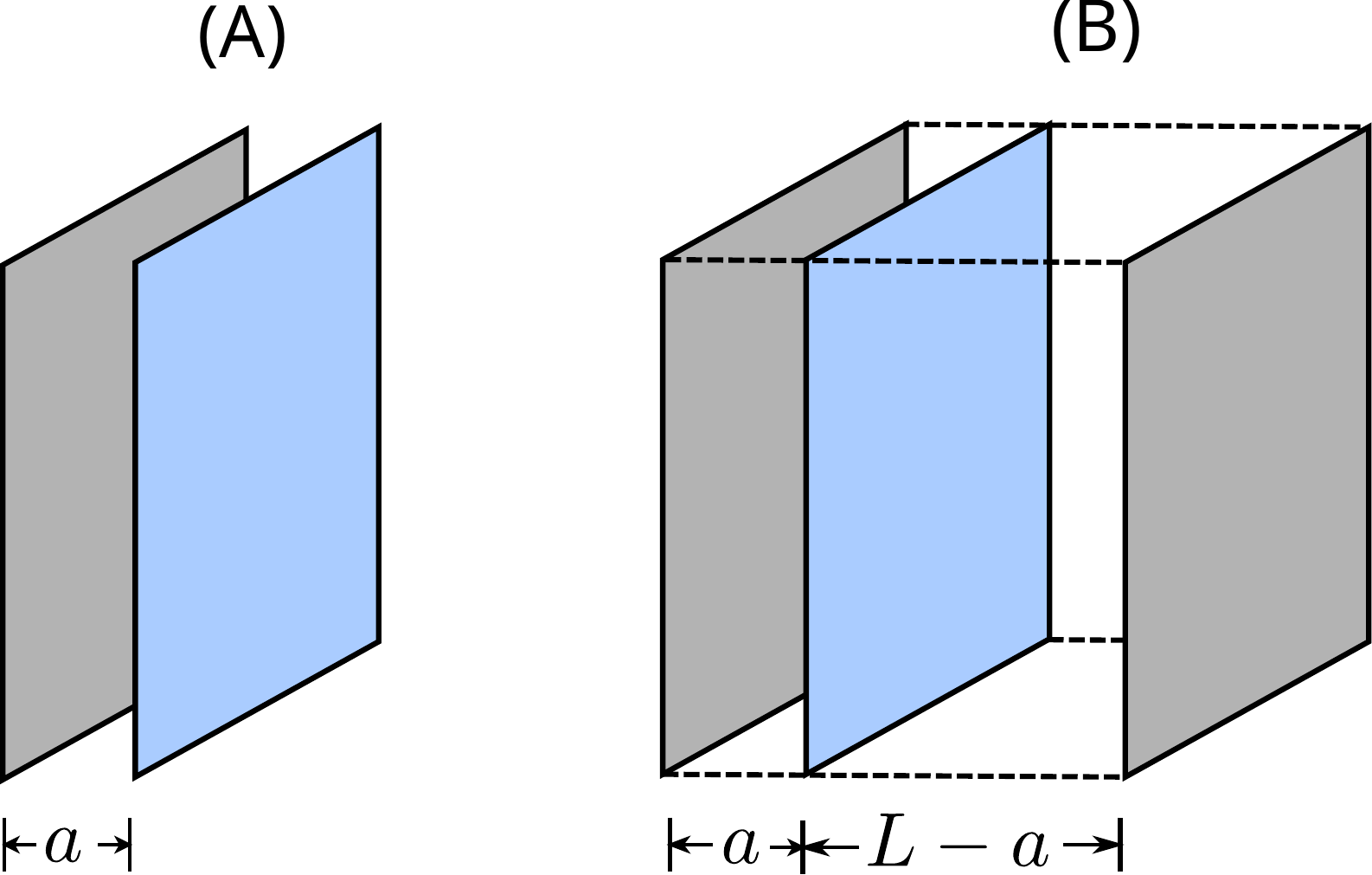}
    \caption{System configuration (A) in condensed derivation of the Casimir force and (B) in Casimir's original paper. In (B) besides the two plates at distance $a$, the enclosing quantization cube of dimension $L$ is considered.}
    \label{fig:casimir-setup}
\end{figure}

The condensed derivation of the Casimir effect starts with considering two metal plates of size $L\times L$ placed in parallel at a distance $a$ from each other in the $z$-direction (horizontal); see Figure~\ref{fig:casimir-setup}A.
Along the $x$- and $y$-direction, one also imagines a square container of size $L \times L$ with metal walls.
The reflection on the metal {surfaces} results in the electromagnetic field forming standing waves, with quantum fluctuations present even in the vacuum state.
The electromagnetic standing-wave modes between the reflecting plates are specified by wavevectors
\begin{equation}
 (k_x,k_y,k_z)=\left(n_x \frac{\pi}{L},n_y \frac{\pi}{L},n_z \frac{\pi}{a} \right), 
\end{equation} 
corresponding to frequencies 
\begin{equation}
    \omega(k_x,k_y,k_z)= c \sqrt{k_x^2+k_y^2+k_z^2},
\end{equation}
where $n_x$, $n_y$ and $n_z$ are non-negative integer numbers.
For each of these standing waves, there are generally two transversal polarisation modes.
An exception occurs for $n_z=0$, where only the polarisation perpendicular to the plates satisfies the boundary condition that the electric field parallel to a metal surface must vanish.
While this is also true for the cases $n_x=0$ and $n_y=0$ due to the boundary condition at the metal walls of the container, it would result in only a negligible correction as the standing wavevectors in these directions are dense for large $L$.

In quantum mechanics, each such electromagnetic standing-wave mode can be regarded as a harmonic oscillator.
The vacuum energy $E(a)$ of such a system of harmonic oscillators is the sum over the contribution $\tfrac{1}{2} \hbar \omega (k_x,k_y,k_z)$ for every mode,
\begin{equation}
    E(a) =\sumprime_{n_z=0}^{\infty} \sum^{\infty}_{n_x,n_y=0} \hbar \omega (k_x,k_y,k_z),
    \label{eq:E-a-sum}
\end{equation}
where we use $\sumprime$ to indicate that the term with $n_z=0$ in the sum has a factor $1/2$ due to the above-mentioned absence of one polarisation mode for $n_z=0$.
We have deliberately separated the summations over $n_x$ and $n_y$, which are now to be transformed into integrals assuming $L$ is large.
Indeed, for large $L$, the spacings of standing wavevectors in the $x$- and $y$-directions $\delta k_x= \delta k_y = \pi/L$ are small. As a result, the summation over $n_x$ and $n_y$ can be replaced by integrals over wavevectors $k_x$ and $k_y$, giving
\begin{equation}
    \sum^{\infty}_{n_x,n_y=0} \omega (k_x,k_y,k_z) \delta k_x \delta k_y = \int_{0}^{\infty} {d} k_x \int_{0}^{\infty} {d} k_y \omega (k_x,k_y,k_z).
    \label{eq:integral}
\end{equation}
One then obtains
\begin{widetext}
\begin{equation}
    E(a) = \frac{\hbar c L^2}{\pi^2} \sumprime_{n_z=0}^{\infty} \int_{0}^{\infty} {d} k_x \int_{0}^{\infty} {d} k_y \sqrt{k_x^2+k_y^2+\left(\frac{n_z \pi}{a}\right)^2}.
\label{eq:E-a}
\end{equation}
\end{widetext}
The integral over $k_x$ and $k_y$ in Eq.~\eqref{eq:E-a} can be expressed in polar coordinates, where the integration over the azimuthal angle in the first quadrant yields a factor of $\pi/2$, thus with $\kappa=\sqrt{k_x^2+k_y^2}$,
\begin{equation}
    E(a) = \frac{\hbar c L^2}{\pi^2} \frac{\pi}{2} \sumprime_{n_z=0}^{\infty} \int_0^{\infty} {d} \kappa \kappa \sqrt{\kappa^2+\left(\frac{n_z \pi}{a}\right)^2}.
\label{eq:E-a-polar}
\end{equation}

{The next steps are sometimes compressed to the following}: consider the case where $a$ is large, so that the summation over $n_z$ in Eq.~\eqref{eq:E-a-polar} can also be transformed into an integral over $k_z$ and therefore
\begin{equation}
    E(a \to \infty)= \frac{\hbar c L^2}{2 \pi} \frac{a}{\pi} \int_{0}^{\infty} d k_z \int_{0}^{\infty} {d} \kappa \kappa \sqrt{\kappa^2+k_z^2}.
\label{eq:E-infty}
\end{equation}
Note that although we write $a\to \infty$, Eq.~\eqref{eq:E-infty} still contains $a$ and thus cannot be regarded as a mathematical limit, but rather as an asymptotic expression describing how $E(a)$ behaves at large $a$.
Following {the compressed} argument, one may be tempted to proceed with computing $\tilde U(a) = E(a) - E(a \to \infty)$, interpreted as the energy to bring the plates from a far distance to the distance $a$. By inserting Eq.~\eqref{eq:E-infty} and Eq.~\eqref{eq:E-a-polar}, this would imply
\begin{widetext}
\begin{equation}
    \tilde U(a)= \frac{\hbar c L^2}{2 \pi} \left[ \sumprime_{n_z=0}^{\infty} \int_0^{\infty} {d} \kappa \kappa \sqrt{\kappa^2+\left(\frac{n_z \pi}{a}\right)^2} - \frac{a}{\pi} \int_{0}^{\infty} d k_z \int_{0}^{\infty} {d} \kappa \kappa \sqrt{\kappa^2+k_z^2} \right].
\label{eq:E-a-polar-subtraction}
\end{equation}
\end{widetext}
This expression is then subject to regularisation (to be discussed below) and the expression for the Casimir force is obtained via $F=\tilde{U}'(a)$.
However, a compressed derivation of this form is inconsistent. Indeed, the symbol $a$ in Eq.~\eqref{eq:E-a-polar} and in Eq.~\eqref{eq:E-infty} refers to different distances: in $E(a)$ it is considered to be small, while in $E(a\to\infty)$ it is asymptotically large.
It just so happens that denoting these two different quantities by the same symbol brings $\tilde U(a)$ into the same {algebraic form} as the correct Casimir interaction energy $U(a)$ (see below) and then also yields the correct expression for the Casimir force.

Let us now return to the original derivation by Casimir \cite{Casimir1948}. Casimir considered a somewhat different setup: the two above plates themselves are contained in a cube of dimension $L$; see Figure~\ref{fig:casimir-setup}B.
This is also similar to the setup considered in, for example, Ref.~\cite{Schwartz2013a}.
The zero-point energy is then given by the sum of the left and the right compartments in Figure~\ref{fig:casimir-setup}B,
\begin{equation}
    E_{\mathrm{tot}}(a)= E_{\mathrm{left}} (a) + E_{\mathrm{right}}(a).
\end{equation}

The zero-point energy for the left compartment $E_{\mathrm{left}}(a)$ can be calculated as before, resulting in the expression $E_{\mathrm{left}}(a) = E(a)$ as given in Eq.~\eqref{eq:E-a-polar}, and analogously $E_{\mathrm{right}} (a) = E(L-a)$.
We note that for large $L$ we can replace the sum by an integral,
\begin{equation}
    E(L-a) = \frac{\hbar c L^2}{2 \pi} \frac{L-a}{\pi} \int_{0}^{\infty} d k_z \int_{0}^{\infty} {d} \kappa \kappa \sqrt{\kappa^2+k_z^2},
    \label{eq:E(L-a)}
\end{equation}
yielding the total energy
\begin{equation}
    E_{\mathrm{tot}}(a) = \frac{\hbar c L^2}{2 \pi} \sumprime_{n_z=0}^{\infty} \int_0^{\infty} {d} \kappa \kappa \sqrt{\kappa^2+\left(\frac{n_z \pi}{a}\right)^2} +\frac{\hbar c L^2}{2 \pi} \frac{L-a}{\pi} \int_{0}^{\infty} d k_z \int_{0}^{\infty} {d} \kappa \kappa \sqrt{\kappa^2+k_z^2} .
\end{equation}
For large $L$, this expression contains an `infinity' independent of $a$ in the last term. This `infinity' can be simply removed by redefining the energy reference point, resulting in what Casimir called interaction energy, $U (a) = E_{\mathrm{tot}}(a) - E_{\mathrm{tot}}(L/2)$. As for large $a$ also the summation in $E_{\mathrm{left}}(a)=E(a)$ can be replaced by an integral as that for $E(L-a)$, giving
\begin{equation}
    E_{\mathrm{tot}}(L/2) = \frac{\hbar c L^2}{2 \pi} \frac{L}{\pi} \int_{0}^{\infty} d k_z \int_{0}^{\infty} {d} \kappa \kappa \sqrt{\kappa^2+k_z^2}.
\end{equation}
The correct expression for the interaction energy is then obtained as
\begin{equation}
    U (a) = \frac{\hbar c L^2}{2 \pi} \left[ \sumprime_{n_z=0}^{\infty} \int_0^{\infty} {d} \kappa \kappa \sqrt{\kappa^2+\left(\frac{n_z \pi}{a}\right)^2} - \frac{a}{\pi} \int_{0}^{\infty} d k_z \int_{0}^{\infty} {d} \kappa \kappa \sqrt{\kappa^2+k_z^2} \right].
    \label{eq:Casimir-interaction-energy}
\end{equation}
While having the same appearance as the expression for $\tilde U(a)$ in Eq.~\eqref{eq:E-a-polar-subtraction}, the quantity $a$ in $E_\mathrm{left}(a)$ and $E_\mathrm{right}(a)$, and therefore in Eq.~\eqref{eq:Casimir-interaction-energy} remains the same (and small).

Let us pause to comment on a slightly different, but also  correct argument that can be found in the literature to arrive at Eq.~\eqref{eq:Casimir-interaction-energy}, for example in Ref.~\cite{Bordag2001New}.
Notice that the energy in the right compartment, $E(L-a)$, is simply its volume $L^2 \times (L-a)$ multiplied by the energy density of the `free vacuum', because the boundary conditions can be ignored for this large volume.
One can interpret that, due to the presence of the plate at distance $a$, a volume of $L^2 \times a$ of the free vacuum (originally of size $L^3$) has been replaced by a modified vacuum.
It is then argued that if one considers only the left compartment (as in Figure~\ref{fig:casimir-setup}A), from $E_\mathrm{left}(a)$ one should subtract the energy of the free vacuum originally contained in the volume $L^2 \times a$ of the left compartment.
This argument also gives directly Eq.~\eqref{eq:Casimir-interaction-energy}, which should not be confused with the one we mentioned at the beginning of this note.
While this argument is consistent, we find Casimir's original one more intuitive.


Coming back to the Casimir interaction energy $U(a)$ in Eq.~\eqref{eq:Casimir-interaction-energy}, it contains the difference of two diverging quantities and hence it is still mathematically \emph{indefinite}. To further proceed, the  interaction energy $U(a)$ must be regularised. For completeness, we reproduce the original regularisation procedure \cite{Casimir1948} here, which we also find very instructive.
Casimir argued that in realistic scenarios, electromagnetic waves of very short wavelengths would not be reflected by the metal plates.
Therefore, in computing the zero-point energy, modes with wavevectors exceeding a certain cutoff $\Lambda$ should be excluded.
This would directly imply that $E(a)$ already in the definition in Eq.~\eqref{eq:E-a-sum} is a sum over a finite number of modes, and is thus finite.
As a result, the Casimir interaction energy in Eq.~\eqref{eq:Casimir-interaction-energy} is well-defined as the difference of two finite numbers.
The result, of course, would depend on the cutoff $\Lambda$.
However, if one assumes that $\Lambda$ is sufficiently large in comparison to the relevant scale of the problem ($1/a$), one can formally consider the limit $\Lambda \to \infty$, which happens to be well-defined. In that way, the cutoff-dependency is removed from the result.

To model this cutoff, one can introduce a regularisation function $f[\omega/(c \Lambda)]$ into Eq.~\eqref{eq:E-a-sum} to obtain the regularised vacuum energy
\begin{equation}
    E_\mathrm{reg}(a) = \sumprime_{n_z=0}^{\infty} \sum^{\infty}_{n_x,n_y=0} f\left[\omega (k_x,k_y,k_z)/(c\Lambda)\right] \hbar \omega (k_x,k_y,k_z).
    \label{eq:E-a-sum-regularised}
\end{equation}
The regularisation function $f$ is $1$ for low frequencies, $\omega <c\Lambda$, but vanishes sufficiently rapidly for the high frequencies, $\omega > c\Lambda$, such that the sum is finite.
Other details of $f$ and the exact value of the cutoff $\Lambda$ are irrelevant; these details are eliminated by assuming that $\Lambda$ is sufficiently large so that the limit $\Lambda \to \infty$ can be taken after evaluating the regularised Casimir interaction energy. Starting from the regularised energy in Eq.~\eqref{eq:E-a-sum-regularised} and following the same computational steps as above, one arrives at the regularised Casimir interaction energy Eq.~\eqref{eq:Casimir-interaction-energy} as
\begin{multline}
    U_\mathrm{reg}(a) = \frac{\hbar c L^2}{2 \pi} \left\{ \sumprime_{n_z=0}^{\infty} \int_0^{\infty} {d} \kappa f \left[ \sqrt{\kappa^2+\left(\frac{n_z \pi}{a}\right)^2}\Big/\Lambda \right] \kappa \sqrt{\kappa^2+ \left(\frac{n_z \pi}{a}\right)^2} \right. \\
    \left.- \frac{a}{\pi} \int_{0}^{\infty} d k_z \int_{0}^{\infty} {d} \kappa f \left[\sqrt{\kappa^2+k_z^2}/\Lambda \right] \kappa \sqrt{\kappa^2+k_z^2} \right\}.
    \label{eq:Casimir-interaction-energy-regularised}
\end{multline}

As $a$ is the natural length scale of the problem, one can introduce the dimensionless integral variables $u=(\kappa a/\pi)^2$ and $x= k_z a/\pi$, yielding
\begin{multline}
    U_\mathrm{reg}(a) = \frac{\hbar c L^2}{4 \pi} \frac{\pi^3}{ a^3} \left\{ \sumprime_{n_z=0}^{\infty} \int_0^{\infty} {d} u f\left[ \sqrt{u+n_z^2}\, \frac{\pi}{a \Lambda}\right] \sqrt{u+n_z^2} \right. \\\left. - \int_{0}^{\infty} d x \int_{0}^{\infty} {d} u f\left[ \sqrt{u+x^2} \,\frac{\pi}{a \Lambda}\right] \sqrt{u+x^2} \right\}.
    \label{eq:Casimir-regularisation}
\end{multline}
We now take the limit of large $\Lambda$, making use of the assumed behaviour of the regularisation function $f$. To this end, one defines the auxiliary function
\begin{equation}
    F(x) = \int_0^{\infty} {d} u f\left[ \sqrt{u+x^2}\, \frac{\pi}{a \Lambda}\right] \sqrt{u+x^2}
    \label{eq:F-defined}
\end{equation}
and rewrites the regularised Casimir interaction energy as
\begin{equation}
    U_\mathrm{reg} (a) = \frac{\hbar c L^2}{4 \pi} \frac{\pi^3}{ a^3} \left[ \sumprime_{n_z=0}^{\infty} F(n_z) - \int_{0}^{\infty} d x F(x) \right].
\end{equation}
One then applies the Euler--Maclaurin formula
\begin{equation}
    \sumprime_{n_z=0}^{\infty} F(n_z) - \int_{0}^{\infty} d x F(x) = - \frac{1}{12} F'(0) + \frac{1}{24 \times 30} F'''(0)+ \cdots
\end{equation}
To compute $F'(x)$, one rewrites Eq.~\eqref{eq:F-defined} as $F(x)= 2 \int_{x}^{\infty} {d} w f[ w \pi/(a \Lambda)] w^2$ using a new integral variable $w=\sqrt{u+x^2}$. It follows\footnote{In his original paper, Casimir used a different variable $w'=u+x^2$ and obtained $F(x)= \int_{x^{2}}^{\infty} {d} w' f[ \sqrt{w'} \pi/(a \Lambda)] \sqrt{w'}$, which was printed as $F(x)= \int_{x^{2}}^{\infty} {d} w' f[ w' \pi/(a \Lambda)] \sqrt{w'}$. This appears to be a simple typographical error and subsequent expressions remain correct.} then $F'(x) = - 2 x^2 f[x \pi/(a \Lambda)]$. Therefore $F'(0)=0$, $F'''(0)= -4$. Higher derivatives contain powers of $\pi/(a \Lambda)$ and therefore vanish in the limit $a \Lambda \to \infty$. Thus one obtains
\begin{equation}
    U_\mathrm{reg}(a) = - \frac{\hbar c L^2}{a^3} \frac{\pi^2}{24 \times 30}.
    \label{eq:Casimir-energy-final}
\end{equation}
The expression for the Casimir force is then directly obtained by taking the derivative of the Casimir interaction energy $U_\mathrm{reg}(a)$ with respect to $a$.

\begin{acknowledgements}
We acknowledge questions from the members of the Theoretical Quantum Optics Group at the University of Siegen during our discussions.
In particular, we are thankful to Otfried Gühne and Stefan Nimmrichter for their insightful comments.
This work was supported by the Deutsche Forschungsgemeinschaft (DFG, German Research Foundation, project number 563437167), the
Sino-German Center for Research Promotion
(Project M-0294), the German Federal Ministry of Research, Technology and Space (Project QuKuK, Grant No.~16KIS1618K and Project BeRyQC, Grant No.~13N17292) and the Project EIN Quantum NRW.
\end{acknowledgements}

\bibliography{casimir}
\end{document}